\RequirePackage[orthodox, l2tabu]{nag}
\documentclass[12pt]{iopart}
\usepackage{iopams}
\expandafter\let\csname equation*\endcsname\relax
\expandafter\let\csname endequation*\endcsname\relax
\usepackage{amsmath}
\usepackage{siunitx}
\usepackage{graphicx}
\usepackage{epstopdf}
\usepackage{microtype}
\usepackage[dvipsnames]{xcolor}
\usepackage{hyperref}
\usepackage[figuresright]{rotating}
\begin{document}
\title[]{Bayesian modelling of multiple plasma diagnostics at Wendelstein 7-X}

\author{Sehyun~Kwak$^{1,2}$, J.~Svensson$^{2}$, S.~Bozhenkov$^{2}$, H.~Trimino~Mora$^{2}$, U.~Hoefel$^{2}$, A.~Pavone$^{2}$, M.~Krychowiak$^{2}$, A.~Langenberg$^{2}$, Y.-c.~Ghim$^{1}$ and W7-X~Team}

\address{$^{1}$Department of Nuclear and Quantum Engineering, KAIST, Daejeon 34141, Korea,~Republic~of}
\address{$^{2}$Max-Planck-Institut f\"{u}r Plasmaphysik, 17491 Greifswald, Germany}

\ead{sehyun.kwak@ipp.mpg.de}

\vspace{10pt}
\begin{indented}
\item[]\today
\end{indented}

\begin{abstract}
	Consistent inference of the electron density and temperature has been carried out with multiple heterogeneous plasma diagnostic data sets at Wendelstein 7-X. The predictive models of the interferometer, Thomson scattering and helium beam emission spectroscopy systems have been developed in the Minerva framework and combined to a single joint model. The electron density and temperature profiles are modelled by Gaussian processes with their hyperparameters. The model parameters such as the calibration factor of the Thomson scattering system and the model predictive uncertainties are regarded as additional unknown parameters. The joint posterior probability distribution of the electron density and temperature profiles, hyperparameters of the Gaussian processes and model parameters is explored by Markov chain Monte Carlo algorithms. The posterior samples drawn from the joint posterior distribution are numerically marginalised over the hyperparameters and model parameters to obtain the marginal posterior distributions of the electron density and temperature profiles. The inference of these profiles is performed with different combinations of the interferometer and Thomson scattering data as well as either the empirical electron density and temperature constraints at the limiter/divertor positions introduced by \textit{virtual observations} or the edge density and temperature from the helium beam emission data. Furthermore, the addition of the X-ray imaging crystal spectrometers to the joint model for the ion temperature profiles is demonstrated. All these profiles presented in this work are inferred with the optimal hyperparameters and model parameters by exploring the full joint posterior distribution which intrinsically embodies Bayesian Occam's razor.
\end{abstract}

%
\vspace{2pc}


%
%

\section{Introduction}\label{sec:introduction}
Consistent inference of physics parameters of fusion plasmas with their associated uncertainties is crucial to understand and to control the underlying physical phenomena in a large scale fusion experiment. Such an experiment as the Joint European Torus (JET) \cite{Litaudon2017} and Wendelstein 7-X (W7-X) \cite{Klinger2019} typically employs several tens of sophisticated and complicated measurement techniques. The analysis of experimental data from each of the measurement instruments is substantially complex, therefore, to make full use of these heterogeneous data sets to refine physics parameters as rigorously as possible is challenging. In order to make this possible and practical, it is advantageous to use a framework that is capable of handling and keeping track of parameters, assumptions, predictive models and observations.

The Minerva framework has been developed to achieve consistent inference for a complex system by modularisation of models and standardisation of interfaces to connect them in a systematic way \cite{Minerva}. For example, a Minerva Thomson scattering (forward) model encapsulates physics and instrumental effects of a Thomson scattering system to calculate Thomson scattering signals given the laser power and wavelength, scattering angles, spectral response functions, data acquisition systems, physics parameters, i.e., electron density and temperature, and so on. These model dependencies can be fed from either other Minerva models or data sources (interfaces to databases), and model predictions such as predicted Thomson scattering signals can be directly compared to the corresponding observations. The Minerva framework automatically manages such integrations of Minerva models, and these Minerva models can be represented by a Bayesian graphical model \cite{Pearl1988}, which is a transparent way of unfolding and handling the complexity of the models. These automatic model administration together with graphical representation make the joint analysis of multiple heterogeneous data sets achievable and practical. In nuclear fusion research, the Minerva framework has been used for a number of scientific applications to magnetic sensors \cite{Svensson2008}, interferometers \cite{Ford2010,Svensson2011_GP}, Thomson scattering systems \cite{Bozhenkov2017,Kwak2020}, soft X-ray spectroscopy \cite{Li2013}, beam emission spectroscopy \cite{Kwak2016,Kwak2017}, X-ray imaging crystal spectrometers \cite{Langenberg2016}, electron cyclotron emission \cite{Hoefel2019} and effective ion charge diagnostics \cite{Pavone2019_Zeff}. These Minerva models can be accelerated by a field-programmable gate array (FPGA) \cite{Mora2017} or an artificial neural network \cite{Pavone2018,Pavone2019}.

In this work, the Bayesian joint analysis of the interferometer, Thomson scattering and helium beam emission spectroscopy systems at W7-X has been carried out (followed by the addition of the X-ray imaging crystal spectrometers \cite{Langenberg2016}). The conventional analysis of the interferometer \cite{Knauer2016}, Thomson scattering \cite{Bozhenkov2017} and helium beam emission spectroscopy \cite{Barbui2016} systems is typically carried out individually. The Thomson scattering system provides the local measurements of the electron density and temperature across the plasma, and the interferometer system measures line integrated electron density along the line of sight. When the calibration factor of the Thomson scattering system has not been fully identified, the electron density profiles from Thomson scattering data can be cross-calibrated with the interferometer data. For this reason, the line of sight of the W7-X interferometer system is set to be approximately identical to the laser path of the Thomson scattering system. In order to perform the cross-calibration as precise as possible, the electron density and temperature in the edge region should be known. However, the Thomson scattering data does not typically provide a good quality of the electron density and temperature measurements in the edge region, where the electronics noise is much larger than the Thomson scattering signals. The conventional way of dealing with this problem is to assume that the electron density and temperature is zero outside the last closed magnetic flux surface (LCFS) given by the variational moments equilibrium code (VMEC) \cite{Hirshman1986,Geiger2010}. The method developed in this work makes use of either the empirical electron density and temperature constraints at the limiter/divertor positions based on physics knowledge \textit{a priori} introduced by \textit{virtual observations} or the edge electron density and temperature measurements from the helium emission spectroscopy system. Furthermore, the X-ray imaging crystal spectrometers (XICS) \cite{Langenberg2016} is integrated to the Bayesian joint model of the interferometer, Thomson scattering and helium beam emission spectroscopy systems in order to infer the electron density and temperature profiles as well as the ion temperature profiles consistent with all these measurements.

\section{The model}\label{sec:bayesian_model}
In Bayesian inference \cite{Pearl1988,Jaynes2003,Sivia2006}, the probability of a hypothetical value of unknown parameters $P\left(H\right)$ can be updated to the \textit{posterior probability} of the unknown parameters given observations $P\left(H|D\right)$ through Bayes formula:
\begin{equation}
P\left(H|D\right)=\frac{P\left(D|H\right)P\left(H\right)}{P\left(D\right)}.
\label{eq:Bayes_formula}
\end{equation}
The probability of the unknown parameters, also known as the prior probability $P\left(H\right)$, encodes the \textit{prior knowledge} such as physics and empirical assumptions. For instance, the temperature must be positive by definition, thus, the probability of any negative temperature must be zero. The conditional probability of the observations $P\left(D|H\right)$ makes a \textit{predictive distribution} over the observations given a hypothetical value of the unknown parameters. In other words, this predictive distribution expresses all possible values of the observations that can be measured given a hypothetical value of the unknown parameters. Typically, the mean of the predictive distribution is given by a function which encapsulates the processes happening during an experiment by taking into account physical phenomena as well as the experimental setup, for example, the instrument effects, calibrations, optics, electronics and so on, also known as a \textit{forward model} $f\left(H\right)$. The marginal probability of the observations $P\left(D\right)$, also known as the \textit{model evidence}, is a normalisation constant in this context.

When we have multiple heterogeneous data sets, which conditionally depend on the unknown parameters, Bayes formula can be written as:
\begin{equation}
P\left(H|\right\{D_{i}\left\}\right)=\frac{\big(\prod_{i}P\left(D_i|H\right)\big)P\left(H\right)}{P\left(\right\{D_{i}\left\}\right)}.
\label{eq:Bayes_formula_heterogenous}
\end{equation}
Each of the predictive distributions contains a forward model of the measurement instruments, which are typically sophisticated and complicated. They may include extra model parameters, for instance calibration factors, depending on the experimental setup. The prior distribution encodes the prior knowledge of the unknown parameters as well as hyperparameters (parameters of the prior distribution) and unknown model parameters. These prior and predictive distributions together constitute the joint probability distribution $P\left(H,D\right)$ which embodies the full relationship between all the unknown parameters and observations. This joint distribution is modelled in Minerva as a Bayesian graphical model \cite{Pearl1988}.

\begin{sidewaysfigure}
	\includegraphics[width=\linewidth]{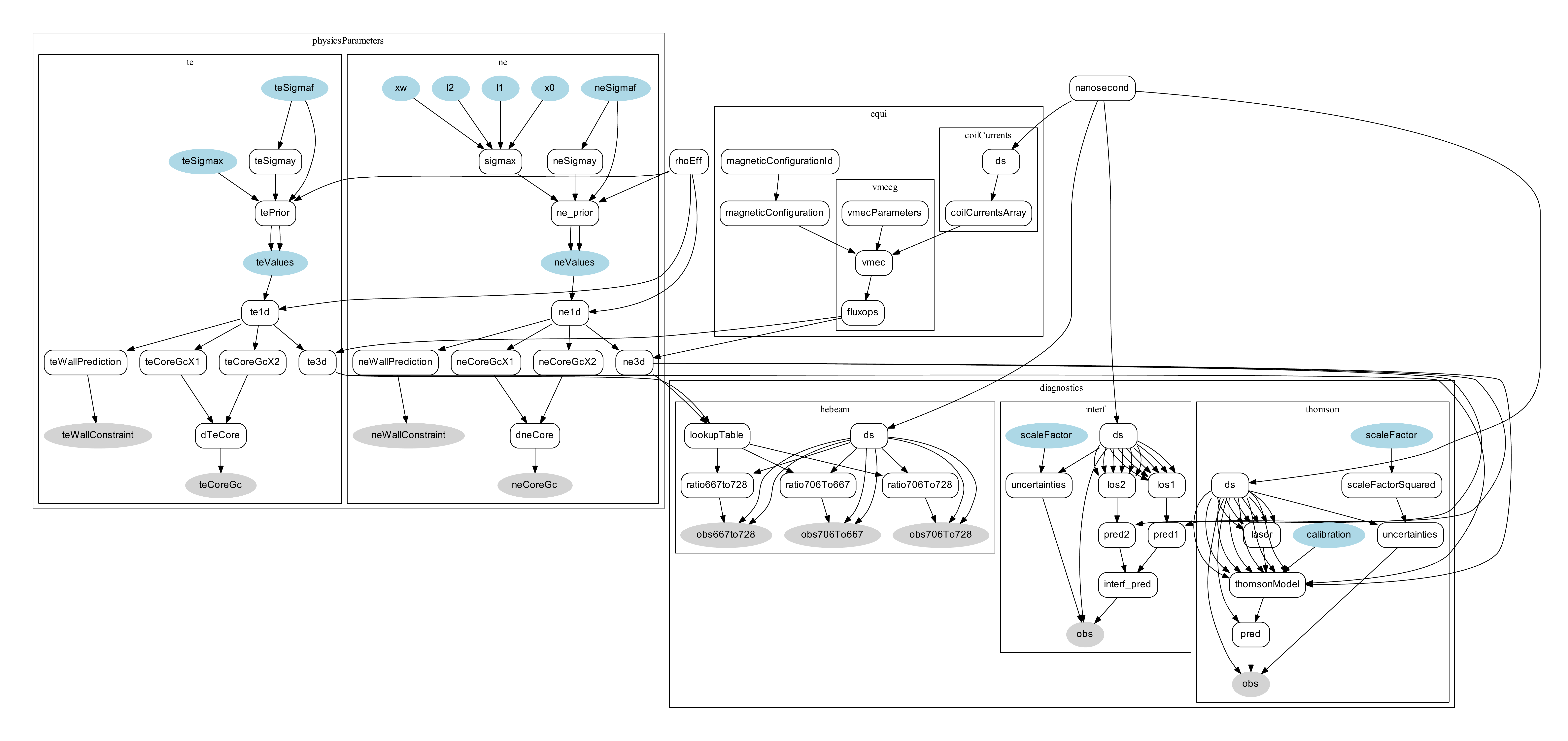}
	\caption{The Minerva graph of the Bayesian joint model of the interferometer, Thomson scattering and helium beam emission spectroscopy systems at Wendelstein 7-X. The unknown parameters and observations are shown as the blue and grey circles, respectively. The electron density $n_\mathrm{e}$ and temperature $T_\mathrm{e}$ are given as a function of the effective minor radius $\rho_\mathrm{eff}$ and mapped to $x,y,z$ Cartesian coordinates through the coordinate transformations provided by the variational moments equilibrium code (VMEC) node. The electron density and temperature profiles are modelled by Gaussian processes with their hyperparameters, and each of the model predictions is calculated given all these unknown parameters. This graph represents the joint probability of all the unknown parameters and observations which consists of all these prior and predictive distributions.}
	\label{fig:model}
\end{sidewaysfigure}

The Minerva graph of the Bayesian joint model of the interferometer, Thomson scattering and helium beam emission spectroscopy systems is shown in Figure~\ref{fig:model}. Each node represents either a deterministic calculation (white box) or a probability function, a prior (blue circle) or a predictive probability (grey circle). Such deterministic nodes include a simple operation (e.g. \texttt{los}, a function for line integration along a line of sight), a physics model (e.g. \texttt{Thomson model}) and a data source (\texttt{ds}). The arrows indicate the conditional dependencies of these nodes. This graph represents the joint distribution of all the unknown parameters and observations which consists of all these prior and predictive distributions.

In this work, the electron density $n_\mathrm{e}$ and temperature $T_\mathrm{e}$ profiles are given as a function of the effective minor radius $\rho_\mathrm{eff}$ and modelled by Gaussian processes \cite{OHagan1978,Neal1995,Rasmussen2006}. The Gaussian processes are \textit{non-parametric} functions which associate any set of points in the domain of the functions with a random vector following a multivariate Gaussian distribution. The properties of the Gaussian processes are determined not by any parametric form but by the covariance function of the Gaussian distribution. The covariance function provides the covariance value between any two points, and the smoothness of the Gaussian processes is determined by these covariance values. In nuclear fusion research, Gaussian processes were first introduced by non-parametric tomography of the electron density and current distribution \cite{Svensson2011_GP}, and has since been used in a number of applications \cite{Kwak2020,Li2013,Kwak2016,Kwak2017,Langenberg2016,Chilenski2015}.

The prior distribution of the electron temperature is given by a Gaussian process with the zero mean and squared exponential covariance function, which is one of the most common specifications of the Gaussian processes, which can be written as:
\begin{align}
P\left(T_\mathrm{e}|\sigma_{T_\mathrm{e}}\right)&=\mathcal{N}\left(\mu_{T_\mathrm{e}},\Sigma_{T_\mathrm{e}}\right),\label{eq:Te_prior}\\
\mu_{T_\mathrm{e}}\left(\rho_\mathrm{eff}\right)&=0,\label{eq:Te_mean}\\
\Sigma_{T_\mathrm{e}}\left(\rho_{\mathrm{eff},i},\rho_{\mathrm{eff},j}\right)&=\sigma_{f,T_\mathrm{e}}^2\exp{\left(-\frac{\left(\rho_{\mathrm{eff},i}-\rho_{\mathrm{eff},j}\right)^2}{2\sigma_{x,T_\mathrm{e}}^2}\right)}+\sigma_{y,T_\mathrm{e}}^2\delta_{ij}.\label{eq:Te_covariance}
\end{align}
All the hyperparameters are denoted as $\sigma_{T_\mathrm{e}}=\left[\sigma_{f,T_\mathrm{e}},\sigma_{x,T_\mathrm{e}}\right]$, and $\sigma_{y,T_\mathrm{e}}$ is set to be relatively small number with respect to $\sigma_{f,T_\mathrm{e}}$, for example $\sigma_{y,T_\mathrm{e}}/\sigma_{f,T_\mathrm{e}}=10^{-3}$ to avoid numerical instabilities. The electron density can have substantially different smoothness (gradient) in the core and edge regions, and for this reason, the prior distribution of the electron density is modelled by a Gaussian process with the zero mean and non-stationary covariance function \cite{Higdon1999} which can be written as:
\begin{align}
P\left(n_\mathrm{e}|\sigma_{n_\mathrm{e}}\right)&=\mathcal{N}\left(\mu_{n_\mathrm{e}},\Sigma_{n_\mathrm{e}}\right),\label{eq:ne_prior}\\
\mu_{n_\mathrm{e}}\left(\rho_\mathrm{eff}\right)&=0,\label{eq:ne_mean}\\
\Sigma_{n_\mathrm{e}}\left(\rho_{\mathrm{eff},i},\rho_{\mathrm{eff},j}\right)&=\sigma_{f,n_\mathrm{e}}^2\left(\frac{2\sigma_{x,n_\mathrm{e}}\left(\rho_{\mathrm{eff},i}\right)\sigma_{x,n_\mathrm{e}}\left(\rho_{\mathrm{eff},j}\right)}{\sigma_{x,n_\mathrm{e}}\left(\rho_{\mathrm{eff},i}\right)^2+\sigma_{x,n_\mathrm{e}}\left(\rho_{\mathrm{eff},j}\right)^2}\right)^{\frac{1}{2}}\nonumber\\
&\times\exp{\left(-\frac{\left(\rho_{\mathrm{eff},i}-\rho_{\mathrm{eff},j}\right)^2}{\sigma_{x,n_\mathrm{e}}\left(\rho_{\mathrm{eff},i}\right)^2+\sigma_{x,n_\mathrm{e}}\left(\rho_{\mathrm{eff},j}\right)^2}\right)}+\sigma_{y,n_\mathrm{e}}^2\delta_{ij}.\label{eq:ne_covariance}
\end{align}
The length scale function $\sigma_{x,n_\mathrm{e}}\left(\rho_\mathrm{eff}\right)$ can be given by a hyperbolic tangent function, developed in \cite{Chilenski2015} and also applied in \cite{Kwak2020}, which is:
\begin{equation}
\sigma_{x,n_\mathrm{e}}\left(\rho_\mathrm{eff}\right)=\frac{\sigma_{x,n_\mathrm{e}}^\mathrm{core}+\sigma_{x,n_\mathrm{e}}^\mathrm{edge}}{2}-\frac{\sigma_{x,n_\mathrm{e}}^\mathrm{core}-\sigma_{x,n_\mathrm{e}}^\mathrm{edge}}{2}\tanh{\frac{\rho_\mathrm{eff}-\rho_{\mathrm{eff},0,n_\mathrm{e}}}{\rho_{\mathrm{eff},\mathrm{w},n_\mathrm{e}}}}.
\label{eq:ne_length_scale}
\end{equation}
Again, all the hyperparameters are denoted as $\sigma_{n_\mathrm{e}}=\left[\sigma_{f,n_\mathrm{e}},\sigma_{x,n_\mathrm{e}}^\mathrm{core},\sigma_{x,n_\mathrm{e}}^\mathrm{edge},\rho_{\mathrm{eff},0,n_\mathrm{e}},\rho_{\mathrm{eff},\mathrm{w},n_\mathrm{e}}\right]$.

The electron density and temperature profiles can be mapped to $x,y,z$ Cartesian coordinates through the coordinate transformations provided by the VMEC node. Given 3D fields of the electron density and temperature in real space, each of predictive distributions of the interferometer, Thomson scattering and helium beam emission spectroscopy data can be calculated. The interferometer system \cite{Knauer2016} is a single chord dispersion interferometer which measures the line integrated electron density along the line of sight. The forward model of the interferometer system predicts the line integral of the electron density, which is directly compared to the measurement stored in the W7-X database. The Thomson scattering system \cite{Bozhenkov2017} collects Thomson scattered spectra from 10 to 79 spatial locations along the laser beam across the centre of the plasma. The physics model of Thomson scattering processes \cite{Naito1993} is implemented in the Thomson scattering model \cite{Bozhenkov2017,Kwak2020} which makes predictions of the Thomson scattered spectra given the electron density and temperature. The calibration factor of the Thomson scattering system has not been yet fully identified, thus the calibration factor is regarded as an additional unknown parameter. The interferometer system is designed to cross-calibrate the Thomson scattering system, and for this reason, the line of sight of the W7-X interferometer is set to be approximately identical to the laser path of the Thomson scattering system. 

The joint posterior distribution of the Bayesian joint model of the interferometer and Thomson scattering systems can be written as:
\begin{align}
&P\left(n_\mathrm{e},T_\mathrm{e},\sigma_{n_\mathrm{e}},\sigma_{T_\mathrm{e}},\sigma_\mathrm{DI},\sigma_\mathrm{TS},C_\mathrm{TS}|D_\mathrm{DI},D_\mathrm{TS}\right)\nonumber\\
&=\frac{P\left(D_\mathrm{DI},D_\mathrm{TS}|n_\mathrm{e},T_\mathrm{e},\sigma_{n_\mathrm{e}},\sigma_{T_\mathrm{e}},\sigma_\mathrm{DI},\sigma_\mathrm{TS},C_\mathrm{TS}\right)P\left(n_\mathrm{e},T_\mathrm{e},\sigma_{n_\mathrm{e}},\sigma_{T_\mathrm{e}},\sigma_\mathrm{DI},\sigma_\mathrm{TS},C_\mathrm{TS}\right)}{P\left(D_\mathrm{DI},D_\mathrm{TS}\right)}\nonumber\\
&=\frac{P\left(D_\mathrm{DI}|n_\mathrm{e},\sigma_\mathrm{DI}\right)P\left(D_\mathrm{TS}|n_\mathrm{e},T_\mathrm{e},\sigma_\mathrm{TS},C_\mathrm{TS}\right)P\left(n_\mathrm{e}|\sigma_{n_\mathrm{e}}\right)P\left(T_\mathrm{e}|\sigma_{T_\mathrm{e}}\right)P\left(\sigma_{n_\mathrm{e}}\right)P\left(\sigma_{T_\mathrm{e}}\right)}{P\left(D_\mathrm{DI}\right)P\left(D_\mathrm{TS}\right)}\nonumber\\
&\times P\left(\sigma_\mathrm{DI}\right)P\left(\sigma_\mathrm{TS}\right)P\left(C_\mathrm{TS}\right),\label{eq:posterior}
\end{align}
where $\sigma_{n_\mathrm{e}}$ and $\sigma_{T_\mathrm{e}}$ are the hyperparameters of the Gaussian processes of the electron density and temperature profiles. The predictive distributions $P\left(D_\mathrm{DI}|n_\mathrm{e},\sigma_\mathrm{DI}\right)$ and $P\left(D_\mathrm{TS}|n_\mathrm{e},T_\mathrm{e},\sigma_\mathrm{TS}\right)$ are modelled as Gaussian distributions whose mean and standard deviation are the predictions of the forward models and predictive uncertainties, which are proportional to the measurement uncertainties with scale factors $\sigma_\mathrm{DI}$ and $\sigma_\mathrm{TS}$. These scale factors are regarded as additional unknown parameters due to our incomplete knowledge of measurement uncertainties. These model parameters and the hyperparameters of the Gaussian processes can be optimised to maximise the posterior probability of the model, which takes into account the principle of Occam's razor \cite{Gull1988,MacKay1991}.

The calibration factor of the Thomson scattering system $C_\mathrm{TS}$ is also treated as an additional unknown parameter, therefore, the Thomson scattering system will be automatically cross-calibrated with the interferometer data. Nevertheless, the electron density and temperature in the edge region play an important role in this cross-calibration, since the profile boundary depends on the observations in the edge region. Here, we utilise our physics and empirical knowledge to impose such observations in the edge region by assuming that the electron density and temperature are not significantly high enough to melt down the limiter and divertor of the W7-X experiment \cite{Klinger2019}. These low density and temperature constraints can be introduced by \textit{virtual} observations at the limiter/divertor positions \textit{a~priori} as a part of the prior distributions, which can be written as:
\begin{align}
P\left(D_{\mathrm{v},n_\mathrm{e}}|n_\mathrm{e}\right)=\mathcal{N}\left(n_\mathrm{e}\left(x_\mathrm{wall},y_\mathrm{wall},z_\mathrm{wall}\right),\sigma_{\mathrm{v},n_\mathrm{e}}^2\right),\label{eq:virtual_density}\\
P\left(D_{\mathrm{v},T_\mathrm{e}}|T_\mathrm{e}\right)=\mathcal{N}\left(T_\mathrm{e}\left(x_\mathrm{wall},y_\mathrm{wall},z_\mathrm{wall}\right),\sigma_{\mathrm{v},T_\mathrm{e}}^2\right),\label{eq:virtual_temperature}
\end{align}
where $x_\mathrm{wall}$, $y_\mathrm{wall}$, $z_\mathrm{wall}$ are the spatial locations of the limiter/divertor. The density and temperature constraints at the limiter/divertor are set to be reasonably low: $D_{\mathrm{wall},n_\mathrm{e}}=\SI{e15}{\per\metre\cubed}$, $\sigma_{\mathrm{wall},n_\mathrm{e}}=\SI{e15}{\per\metre\cubed}$, $D_{\mathrm{wall},T_\mathrm{e}}=\SI{0.1}{\electronvolt}$ and $\sigma_{\mathrm{wall},T_\mathrm{e}}=\SI{0.1}{\electronvolt}$. In the same way, we also introduce the zero gradients of the electron density and temperature profiles at the magnetic axis. Given these virtual observations, the joint posterior probability can be written as:
\begin{align}
&P\left(n_\mathrm{e},T_\mathrm{e},\sigma_{n_\mathrm{e}},\sigma_{T_\mathrm{e}},\sigma_\mathrm{DI},\sigma_\mathrm{TS},C_\mathrm{TS}|D_\mathrm{DI},D_\mathrm{TS},D_{\mathrm{v},n_\mathrm{e}},D_{\mathrm{v},T_\mathrm{e}}\right)\nonumber\\
&=\frac{P\left(D_\mathrm{DI},D_\mathrm{TS},D_{\mathrm{v},n_\mathrm{e}},D_{\mathrm{v},T_\mathrm{e}}|n_\mathrm{e},T_\mathrm{e},\sigma_{n_\mathrm{e}},\sigma_{T_\mathrm{e}},\sigma_\mathrm{DI},\sigma_\mathrm{TS},C_\mathrm{TS}\right)P\left(n_\mathrm{e},T_\mathrm{e},\sigma_{n_\mathrm{e}},\sigma_{T_\mathrm{e}},\sigma_\mathrm{DI},\sigma_\mathrm{TS},C_\mathrm{TS}\right)}{P\left(D_\mathrm{DI},D_\mathrm{TS},D_{\mathrm{v},n_\mathrm{e}},D_{\mathrm{v},T_\mathrm{e}}\right)}\nonumber\\
&=\frac{P\left(D_\mathrm{DI}|n_\mathrm{e},\sigma_\mathrm{DI}\right)P\left(D_\mathrm{TS}|n_\mathrm{e},T_\mathrm{e},\sigma_\mathrm{TS},C_\mathrm{TS}\right)P\left(D_{\mathrm{v},n_\mathrm{e}}|n_\mathrm{e}\right)P\left(D_{\mathrm{v},T_\mathrm{e}}|T_\mathrm{e}\right)P\left(n_\mathrm{e}|\sigma_{n_\mathrm{e}}\right)P\left(T_\mathrm{e}|\sigma_{T_\mathrm{e}}\right)}{P\left(D_\mathrm{DI}\right)P\left(D_\mathrm{TS}\right)P\left(D_{\mathrm{v},n_\mathrm{e}}\right)P\left(D_{\mathrm{v},T_\mathrm{e}}\right)}\nonumber\\
&\times P\left(\sigma_{n_\mathrm{e}}\right)P\left(\sigma_{T_\mathrm{e}}\right)P\left(\sigma_\mathrm{DI}\right)P\left(\sigma_\mathrm{TS}\right)P\left(C_\mathrm{TS}\right)\nonumber\\
&=\frac{P\left(D_\mathrm{DI}|n_\mathrm{e},\sigma_\mathrm{DI}\right)P\left(D_\mathrm{TS}|n_\mathrm{e},T_\mathrm{e},\sigma_\mathrm{TS},C_\mathrm{TS}\right)P\left(n_\mathrm{e}|D_{\mathrm{v},n_\mathrm{e}},\sigma_{n_\mathrm{e}}\right)P\left(T_\mathrm{e}|D_{\mathrm{v},T_\mathrm{e}},\sigma_{T_\mathrm{e}}\right)}{P\left(D_\mathrm{DI}\right)P\left(D_\mathrm{TS}\right)}\nonumber\\
&\times P\left(\sigma_{n_\mathrm{e}}\right)P\left(\sigma_{T_\mathrm{e}}\right)P\left(\sigma_\mathrm{DI}\right)P\left(\sigma_\mathrm{TS}\right)P\left(C_\mathrm{TS}\right),\label{eq:constrainted_posterior}
\end{align}
where $P\left(n_\mathrm{e}|D_{\mathrm{v},n_\mathrm{e}},\sigma_{n_\mathrm{e}}\right)$ and $P\left(T_\mathrm{e}|D_{\mathrm{v},T_\mathrm{e}},\sigma_{T_\mathrm{e}}\right)$ are the Gaussian process priors with the edge constraints introduced by these virtual observations. Remarkably, any physics/empirical law can be introduced by virtual observations, for example the left-hand and right-hand side of physics formula can be regarded as predictions and corresponding observations at any space and time. These physics/empirical priors based on virtual observations have been used for the Bayesian joint model at Wendelstein 7-AS \cite{Svensson2004} and the plasma equilibria at JET \cite{Ford2010}.

On the other hand, we can provide local measurements of the electron density and temperature in the edge region from the helium beam emission spectroscopy system. The helium beam emission spectroscopy system \cite{Barbui2016} injects helium gas into the plasma and collects three helium line emissions (\SI{667}{\nano\meter}, \SI{706}{\nano\meter} and \SI{728}{\nano\meter} lines). The electron density and temperature can be inferred from three line intensity ratios of \SI{667}{\nano\meter} to \SI{728}{\nano\meter}, \SI{706}{\nano\meter} to \SI{667}{\nano\meter} and \SI{706}{\nano\meter} to \SI{728}{\nano\meter} helium lines by the pre-calculated lookup tables based on the collisional-radiative model \cite{Kwak2017,Krychowiak2011}. The joint posterior probability of the Bayesian joint model of the interferometer, Thomson scattering and helium beam emission spectroscopy systems can be written as:
\begin{align}
&P\left(n_\mathrm{e},T_\mathrm{e},\sigma_{n_\mathrm{e}},\sigma_{T_\mathrm{e}},\sigma_\mathrm{DI},\sigma_\mathrm{TS},C_\mathrm{TS}|D_\mathrm{DI},D_\mathrm{TS},D_\mathrm{He}\right)\nonumber\\
&=\frac{P\left(D_\mathrm{DI},D_\mathrm{TS},D_\mathrm{He}|n_\mathrm{e},T_\mathrm{e},\sigma_{n_\mathrm{e}},\sigma_{T_\mathrm{e}},\sigma_\mathrm{DI},\sigma_\mathrm{TS},C_\mathrm{TS}\right)P\left(n_\mathrm{e},T_\mathrm{e},\sigma_{n_\mathrm{e}},\sigma_{T_\mathrm{e}},\sigma_\mathrm{DI},\sigma_\mathrm{TS},C_\mathrm{TS}\right)}{P\left(D_\mathrm{DI},D_\mathrm{TS},D_\mathrm{He}\right)}\nonumber\\
&=\frac{P\left(D_\mathrm{DI}|n_\mathrm{e},\sigma_\mathrm{DI}\right)P\left(D_\mathrm{TS}|n_\mathrm{e},T_\mathrm{e},\sigma_\mathrm{TS},C_\mathrm{TS}\right)P\left(D_\mathrm{He}|n_\mathrm{e},T_\mathrm{e}\right)P\left(n_\mathrm{e}|\sigma_{n_\mathrm{e}}\right)P\left(T_\mathrm{e}|\sigma_{T_\mathrm{e}}\right)}{P\left(D_\mathrm{DI}\right)P\left(D_\mathrm{TS}\right)P\left(D_\mathrm{He}\right)}\nonumber\\
&\times P\left(\sigma_{n_\mathrm{e}}\right)P\left(\sigma_{T_\mathrm{e}}\right)P\left(\sigma_\mathrm{DI}\right)P\left(\sigma_\mathrm{TS}\right)P\left(C_\mathrm{TS}\right),\label{eq:full_posterior}
\end{align}
where $D_\mathrm{He}$ is the helium beam emission data. Again, the predictive distribution $P\left(D_\mathrm{He}|n_\mathrm{e},T_\mathrm{e}\right)$ is modelled as a Gaussian distribution whose mean and variance are the predictions of the lookup tables and the predictive uncertainties of these helium line ratios.

All these joint posterior distributions are explored by Markov chain Monte Carlo (MCMC) algorithms, specifically adaptive Metropolis-Hastings algorithms \cite{Metropolis1953,Hastings1970,Haario2001} implemented in Minerva. All the hyperparameters of the Gaussian processes and the model parameters are marginalised out numerically in order to obtain the marginal posterior distributions of the electron density and temperature profiles. This means that these profiles are inferred by taking into account all the possible values of the hyperparameters and model parameters consistent with all the measurements simultaneously.

\section{The inference}\label{sec:inference}
The electron density and temperature profiles are amongst the most important physics parameters to understand magnetohydrodynamic equilibrium, transport and performance of the fusion plasma. The Thomson scattering system provides the electron density and temperature profiles across half of the plasma (upgraded to the full range in the latest campaigns), and the dispersion interferometer system measures the line integrated electron density which can be used to infer the calibration factor and to cross-calibrate the Thomson scattering system since the calibration factor has not been yet fully identified. Nevertheless, the profile boundary plays an important role in this cross-calibration. Since the profile boundary can be determined by the information of the electron density and temperature in the edge region, this information can be provided by either the virtual observations at the limiter/divertor positions or the helium beam emission data. In this work, profile inference has been carried out with different combinations of the interferometer, Thomson scattering systems and helium beam emission data as well as the edge virtual observations.

\begin{figure}
	\centering
	\includegraphics[width=\linewidth]{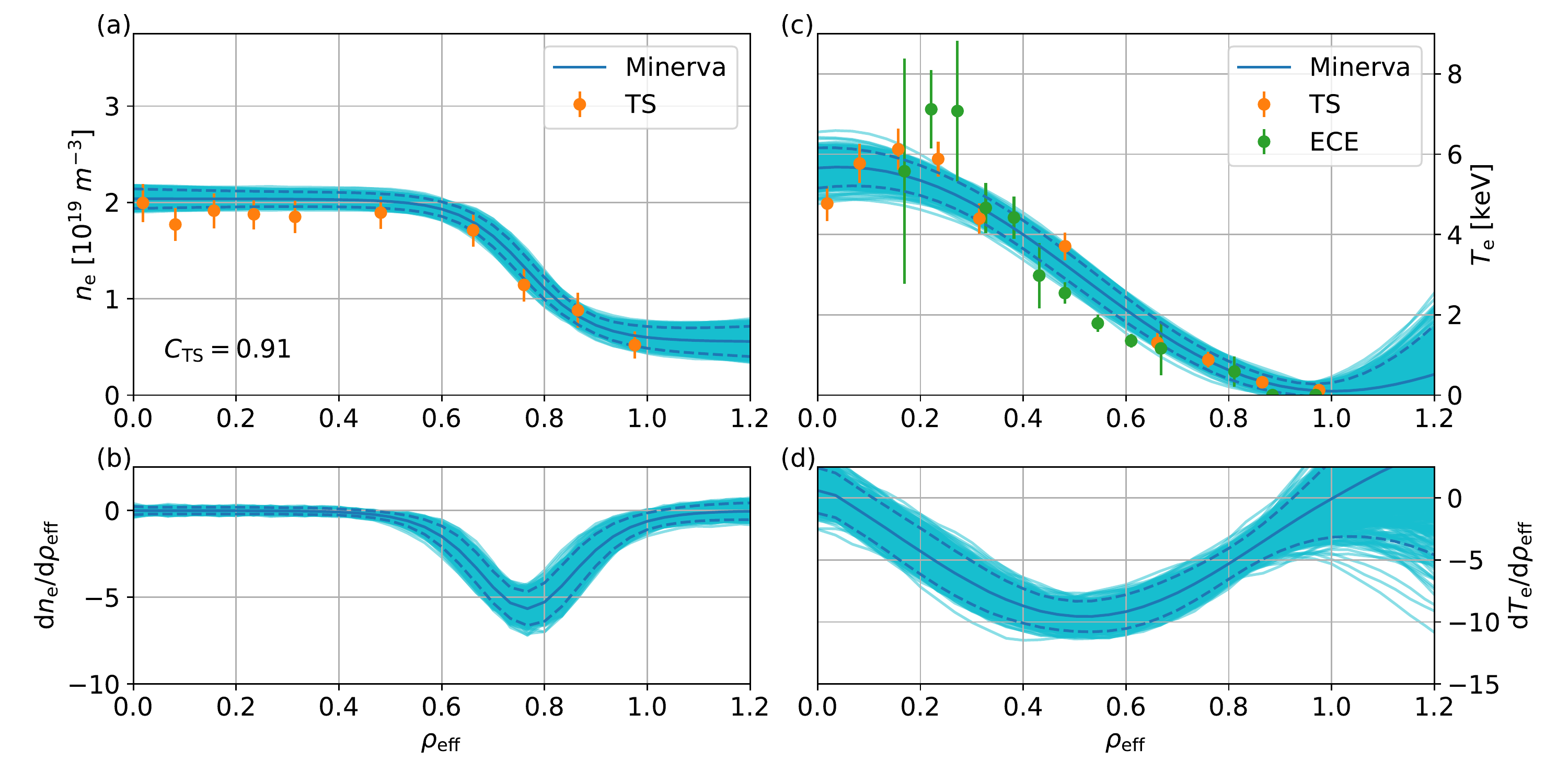}
	\caption{Inference results of the Bayesian joint model of the interferometer and Thomson scattering systems (experiment ID 20160309.013, $t=\SI{0.43}{\second}$): (a) the electron density and (b) temperature profiles and (c) $n_\mathrm{e}$ and (d) $T_\mathrm{e}$ gradient profiles. The blue and light blue lines are the marginal posterior mean and samples, respectively. The orange dots are the electron density and temperature with the error bars provided by the Bayesian Thomson scattering analysis \cite{Bozhenkov2017}. The green dots are the electron temperature from the electron cyclotron emission (ECE) analysis at the low field side \cite{Hirsch2019}. The Thomson scattering system is automatically cross-calibrated with the inferred calibration factor $C_\mathrm{TS}=0.91$ by the joint model. We note that, in this case, the electron density provided by the Thomson scattering system alone (the orange dots) is not consistent with the interferometer data due to some calibration uncertainties \cite{Bozhenkov2017}, whereas the profiles from the joint model (the blue lines) are consistent with both Thomson scattering and interferometer data.}
	\label{fig:profiles}
\end{figure}

Figure~\ref{fig:profiles} shows the electron density and temperature profiles with respect to the effective minor radius $\rho_\mathrm{eff}$ inferred by exploring the joint posterior distribution given the interferometer and Thomson scattering data which is given by Equation~(\ref{eq:posterior}). The blue and light blue lines are the marginal posterior mean and samples, respectively. The marginal posterior samples calculated by numerically integrating the joint posterior distribution over the hyperparameters and model parameters, which can be written as:
\begin{align}
&P\left(n_\mathrm{e},T_\mathrm{e}|D_\mathrm{DI},D_\mathrm{TS}\right)\nonumber\\
&=\int\int\int\int\int P\left(n_\mathrm{e},T_\mathrm{e},\sigma_{n_\mathrm{e}},\sigma_{T_\mathrm{e}},\sigma_\mathrm{DI},\sigma_\mathrm{TS},C_\mathrm{TS}|D_\mathrm{DI},D_\mathrm{TS}\right)\,\mathrm{d}\sigma_{n_\mathrm{e}}\mathrm{d}\sigma_{T_\mathrm{e}}\mathrm{d}\sigma_\mathrm{DI}\mathrm{d}\sigma_\mathrm{TS}\mathrm{d}C_\mathrm{TS}.\label{eq:marginalisation}
\end{align}
The orange dots are the electron density and temperature with the error bars provided by the Thomson scattering analysis implemented in Minerva \cite{Bozhenkov2017}. The green dots are the electron temperature from the electron cyclotron emission (ECE) analysis at the low field side \cite{Hirsch2019}. The calibration factor of the Thomson scattering system is uncertain due to unknown factors during experiments such as laser misalignment, and the electron density profiles of the Thomson scattering analysis, therefore, might not be consistent with the line integrated electron density measurement from the interferometer. On the other hand, the joint model automatically calibrates the Thomson scattering data with the line integrated electron density measurement, thus the electron density profiles of the joint analysis are consistent with both Thomson scattering and interferometer data (the inferred calibration factor $C_\mathrm{TS}=0.91$). In other words, the Thomson scattering analysis might underestimate the electron density profiles by approximately $9\%$ with respect to the interferometer data. The $n_\mathrm{e}$ and $T_\mathrm{e}$ gradient profiles are also presented in Figure~\ref{fig:profiles}(c) and Figure~\ref{fig:profiles}(d). We note that there is no measurement available outside the last closed magnetic flux surface (LCFS), i.e., $\rho_\mathrm{eff}>1.0$ so that the electron density and temperature can be purely determined by the Gaussian process priors.

\begin{figure}
	\centering
	\includegraphics[width=\linewidth]{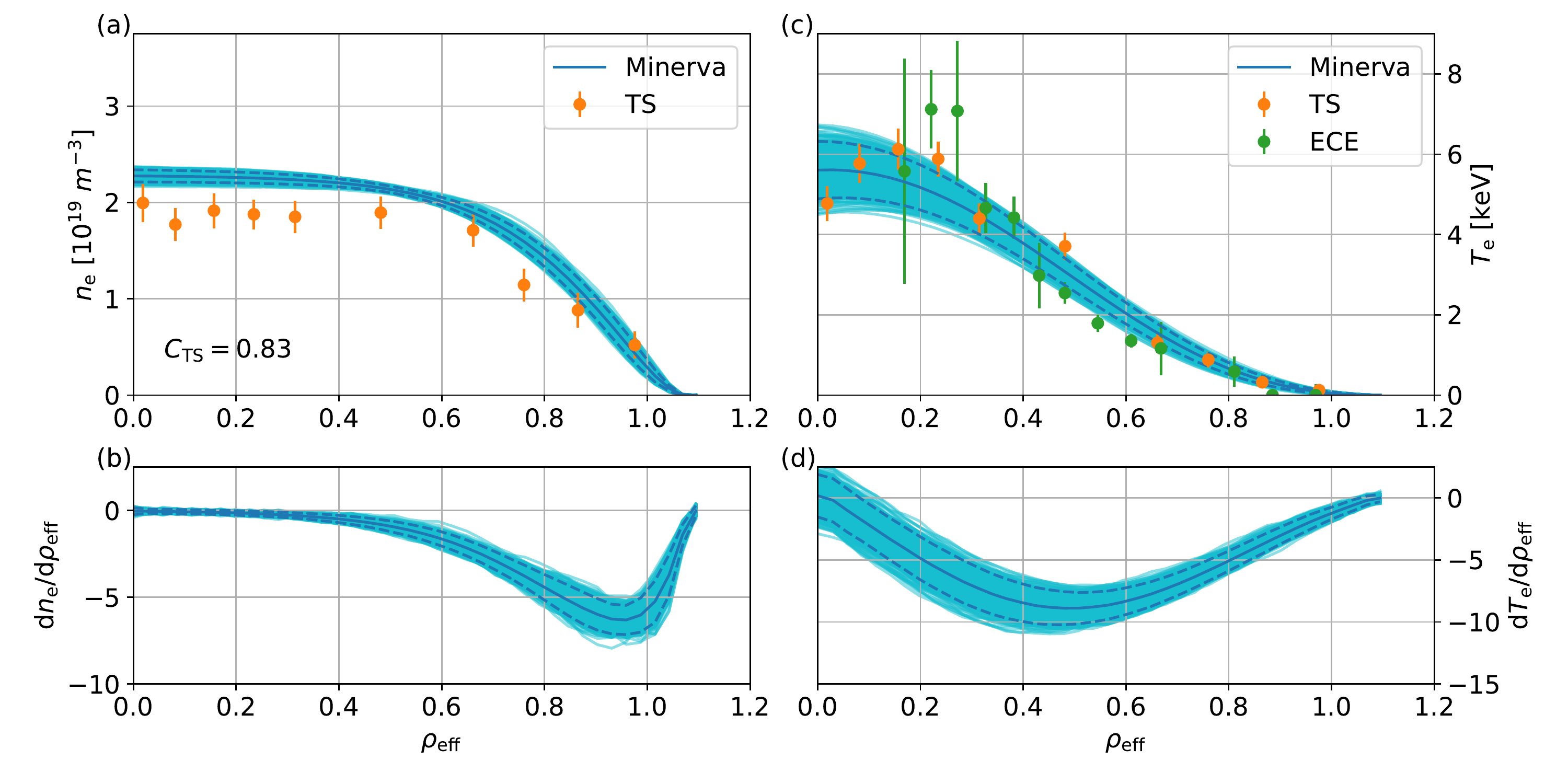}
	\caption{Same as Figure~\ref{fig:profiles} for inference results of the Bayesian joint model of the interferometer and Thomson scattering systems with the electron density and temperature constraints at the limiter/divertor positions introduced by the virtual observations.}
	\label{fig:constrained_profiles}
\end{figure}

The electron density and temperature are not expected to be significantly high at the limiter/divertor positions, and we can introduce this prior knowledge by making the virtual observations, as described in Section~\ref{sec:bayesian_model}. The electron density and temperature profiles of the marginal posterior distribution given these virtual observations $P\left(n_\mathrm{e},T_\mathrm{e}|D_\mathrm{DI},D_\mathrm{TS},D_{\mathrm{v},n_\mathrm{e}},D_{\mathrm{v},T_\mathrm{e}}\right)$ are shown in Figure~\ref{fig:constrained_profiles}. We remark that the mean values of the calibration factor of the Thomson scattering system with and without the virtual observations are substantially different ($C_\mathrm{TS}=0.83$ with the virtual observations and $C_\mathrm{TS}=0.91$ without the virtual observations). In other words, the calibration factor of the Thomson scattering system can be substantially influenced by the information of the electron density and temperature in the edge region.

\begin{figure}
	\centering
	\includegraphics[width=\linewidth]{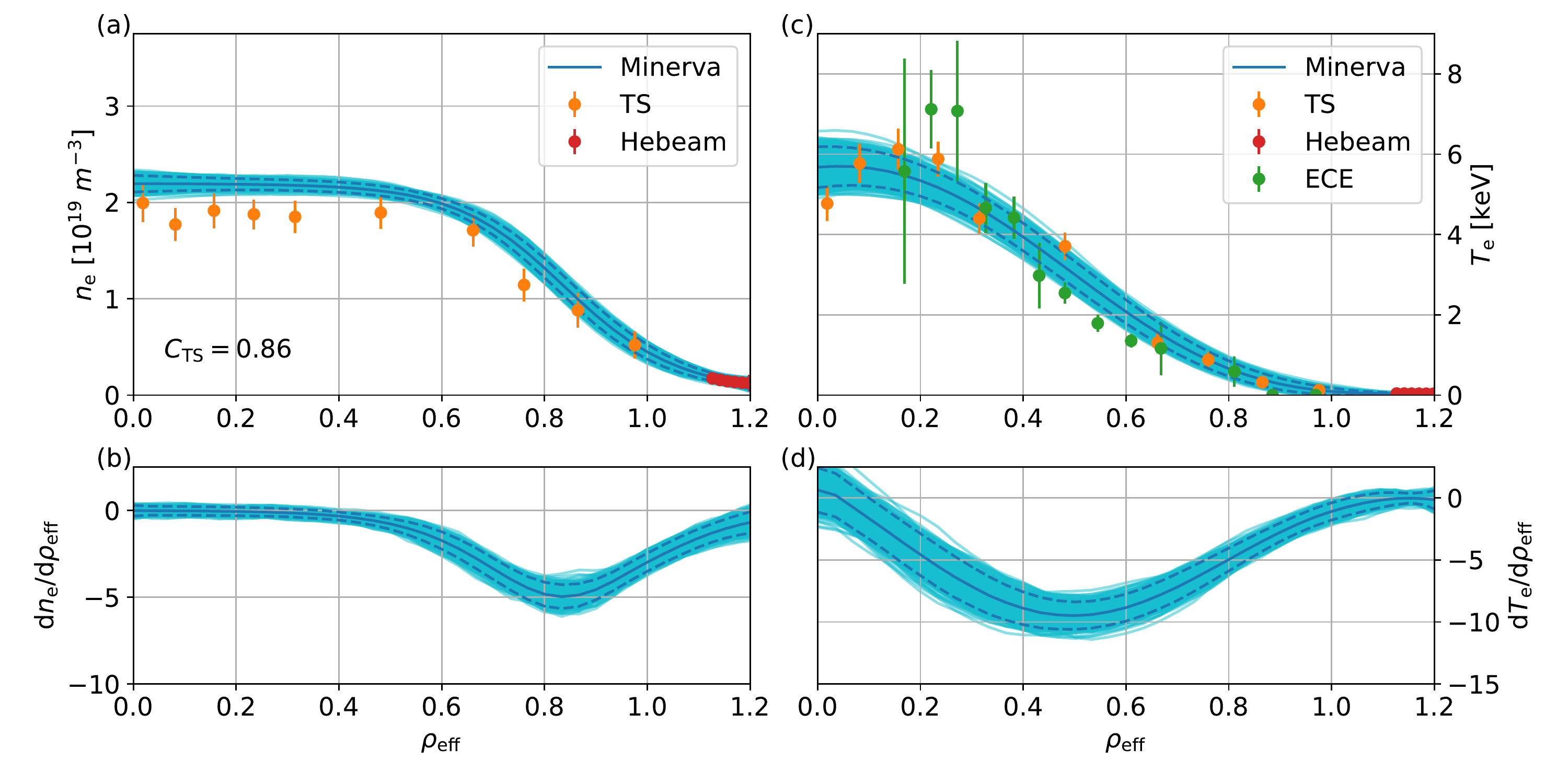}
	\caption{Same as Figure~\ref{fig:profiles} for inference results of the Bayesian joint model of the interferometer, Thomson scattering, and helium beam emission spectroscopy systems. The red dots are the electron density and temperature of the stand-alone analysis of Bayesian helium beam model, developed in this work.}
	\label{fig:hebeam_profiles}
\end{figure}

\begin{figure}
	\centering
	\includegraphics[width=\linewidth]{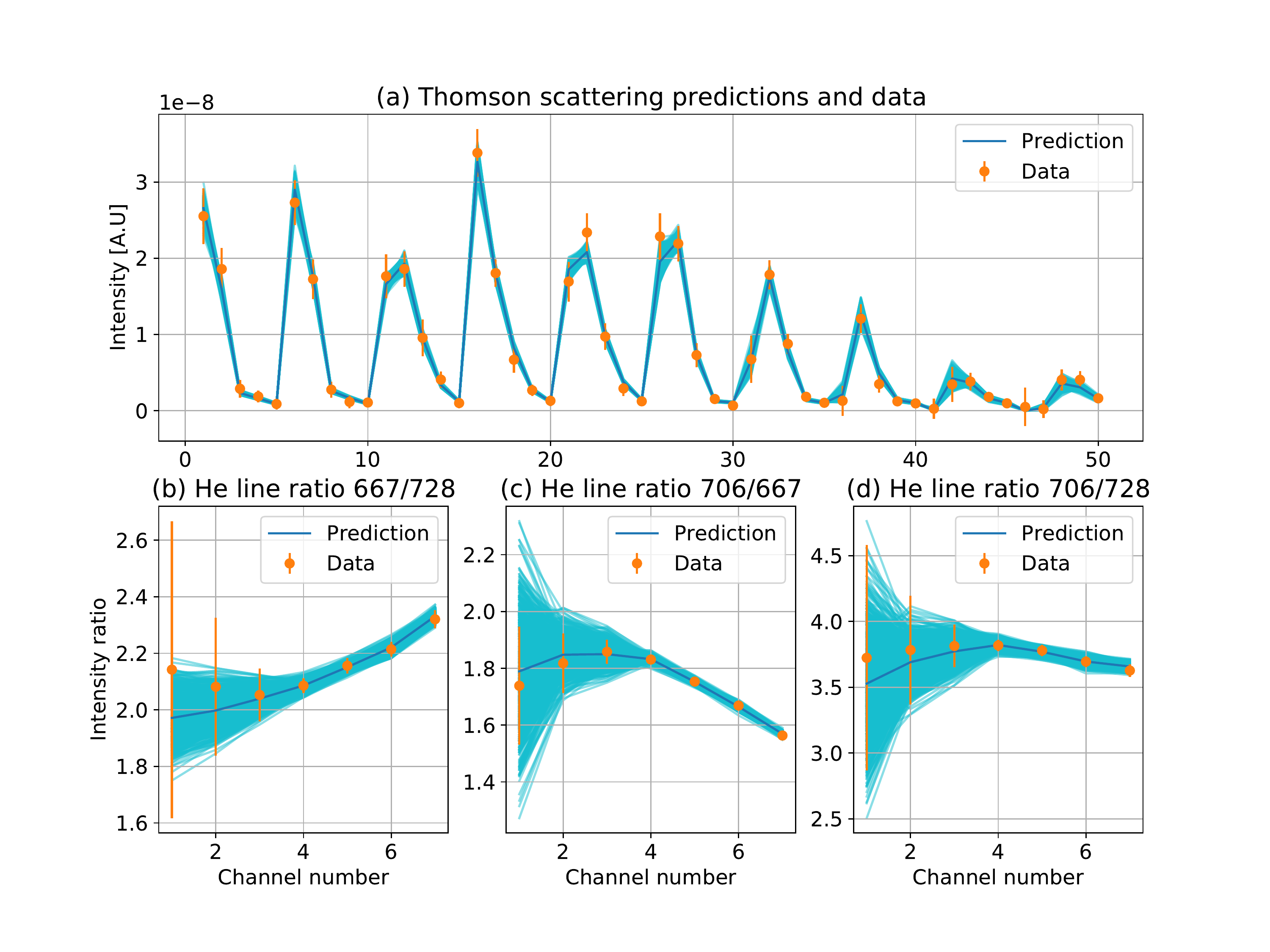}
	\caption{The predictions (in blue and light blue) and observations (in orange) of (a) the Thomson scattering data and (b,c,d) the three helium line intensity ratios given the posterior mean and samples shown in Figure~\ref{fig:hebeam_profiles}. The Thomson scattering signals consists of 50 data points from ten spatial locations (five integrated signals over five different spectral ranges from each spatial location). The helium beam emission data are the three line intensity ratios of (a) \SI{667}{\nano\meter} to \SI{728}{\nano\meter}, (b) \SI{706}{\nano\meter} to \SI{667}{\nano\meter} and (c) \SI{706}{\nano\meter} to \SI{728}{\nano\meter} helium lines from eight spatial locations.}
	\label{fig:predictions}
\end{figure}

In order to compare the inference solutions of the joint model given the virtual and experimental observations in the edge region, the helium beam emission data is added to the joint model instead of the virtual observations. The electron density and temperature profiles of the marginal posterior distribution given the helium beam emission data $P\left(n_\mathrm{e},T_\mathrm{e}|D_\mathrm{DI},D_\mathrm{TS},D_\mathrm{He}\right)$ are shown in Figure~\ref{fig:hebeam_profiles}. The mean value of the calibration factor with the helium beam emission data ($C_\mathrm{TS}=0.86$) is slightly different from the one with the virtual observations ($C_\mathrm{TS}=0.83$). The predictions given these marginal posterior mean and samples and the corresponding observations are compared in Figure~\ref{fig:predictions}. The helium beam emission spectroscopy system provides not only the density and temperature measurements but also their measurement uncertainties in the edge region which are critical to determining the optimal hyperparameters (smoothness) by Bayesian Occam's razor \cite{Gull1988,MacKay1991}. Unlike the inference results given the virtual observations, the joint model of the interferometer, Thomson scattering and helium beam emission spectroscopy systems provides reasonable electron density and temperature profiles in the edge region. Nevertheless, the virtual observations could be another possible option to reinforce the model and exclude physically/empirically improbable solutions when the observations are not sufficiently available.

We emphasise that these profiles neither underfit nor overfit the data. Bayesian methods penalise underfitted and overfitted models automatically and quantitatively. Underfitted models, which propose over-simplified profiles, for example straight profiles, are not able to predict the data within their predictive uncertainties. On the other hand, overfitted model, which propose over-complex profiles, for example wiggly profiles, are able to predict the data better than simpler models. However, overfitted models can propose a greater variety of profiles than simpler models do, and each of them is almost equally probable. The probability of each proposed profile hence is lower than the probability of the profiles proposed by simpler models because the probability over the entire profile space must be equal to one. For this reason, over-complex models are automatically self-penalised by Bayesian Occam's razor \cite{Gull1988,MacKay1991}. In this case, Gaussian processes with too small length scale (over-complex models) are able to propose profiles which predict the data accurately, i.e., high predictive probabilities $P\left(D_\mathrm{DI}|n_\mathrm{e},\sigma_\mathrm{DI}\right)$, $P\left(D_\mathrm{TS}|n_\mathrm{e},T_\mathrm{e},\sigma_\mathrm{TS},C_\mathrm{TS}\right)$ and $P\left(D_\mathrm{He}|n_\mathrm{e},T_\mathrm{e}\right)$, but the prior probabilities of these proposed profiles $P\left(n_\mathrm{e}|\sigma_{n_\mathrm{e}}\right)$ and $P\left(T_\mathrm{e}|\sigma_{T_\mathrm{e}}\right)$ are low since the Gaussian processes can propose many other candidates equally probable. Consequently, the joint posterior probability associated with over-complex models is low. The models with too large predictive uncertainties (over-complex models) are also self-penalised in the same way. By exploring the joint posterior distribution of the electron density and temperature profiles, hyperparameters and model parameters, we collect profiles with proper length scale (smoothness) and predictive uncertainties. Furthermore, these inference solutions provide marginal posterior samples and uncertainties which are obtained by taking into account all possible values of the hyperparameters and model parameters. In other words, these samples and uncertainties do not depend on specific values of hyperparameters and model parameters.

\section{The addition of the X-ray Imaging Crystal Spectrometers}\label{sec:xics_combination}
The X-ray imaging crystal spectrometers (XICS) \cite{Langenberg2016} measure X-ray spectra of argon and iron impurities in different charge states within a wide range of electron temperature, from \SIrange{0.3}{6}{\kilo\electronvolt}. The XICS system collects line integrated spectra along 20 lines of sight, covering more than half of the plasma on the poloidal cross section at a toroidal angle of 159.09. The XICS forward model implemented previously in Minerva \cite{Langenberg2016} is added to the Bayesian joint model of the interferometer, Thomson scattering and helium beam emission spectroscopy systems. The local X-ray spectra are calculated by taking into account a number of atomic processes such as excitation, recombination, ionisation and charge exchange and depend on the electron density and temperature as well as the ion temperature. The forward model integrates these predicted local spectra given these physics parameters along the lines of sight to calculate the line integrated X-ray spectra. 

The ion temperature prior distribution is modelled by a Gaussian process with the zero mean and squared exponential covariance function. The joint posterior probability given the interferometer, Thomson scattering, helium beam emission and XICS data can be written as:
\begin{align}
&P\left(n_\mathrm{e},T_\mathrm{e},T_\mathrm{i},\sigma_{n_\mathrm{e}},\sigma_{T_\mathrm{e}},\sigma_{T_\mathrm{i}},\sigma_\mathrm{DI},\sigma_\mathrm{TS},C_\mathrm{TS}|D_\mathrm{DI},D_\mathrm{TS},D_\mathrm{He},D_\mathrm{XICS}\right)\nonumber\\
&=\frac{P\left(D_\mathrm{XICS}|n_\mathrm{e},T_\mathrm{e},T_\mathrm{i}\right)P\left(T_\mathrm{i}|\sigma_{T_\mathrm{i}}\right)P\left(\sigma_{T_\mathrm{i}}\right)P\left(n_\mathrm{e},T_\mathrm{e},\sigma_{n_\mathrm{e}},\sigma_{T_\mathrm{e}},\sigma_\mathrm{DI},\sigma_\mathrm{TS},C_\mathrm{TS}|D_\mathrm{DI},D_\mathrm{TS},D_\mathrm{He}\right)}{P\left(D_\mathrm{XICS}\right)}
\label{eq:full_posterior_with_xics}
\end{align}
where $T_\mathrm{i}$ is the ion temperature, $\sigma_{T_\mathrm{i}}$ all the hyperparameters of the Gaussian process and $D_\mathrm{XICS}$ the XICS data. The predictive distribution $P\left(D_\mathrm{XICS}|n_\mathrm{e},T_\mathrm{e},T_\mathrm{i}\right)$ is modelled as a Gaussian distribution whose mean and variance are the predictions of the XICS forward model. The predictive uncertainties of the line integrated X-ray spectra. The electron density and temperature profiles as well as the ion temperature profiles are inferred tomographically given the interferometer, Thomson scattering, helium beam emission and XICS data.

\begin{figure}
	\centering
	\includegraphics[width=\linewidth]{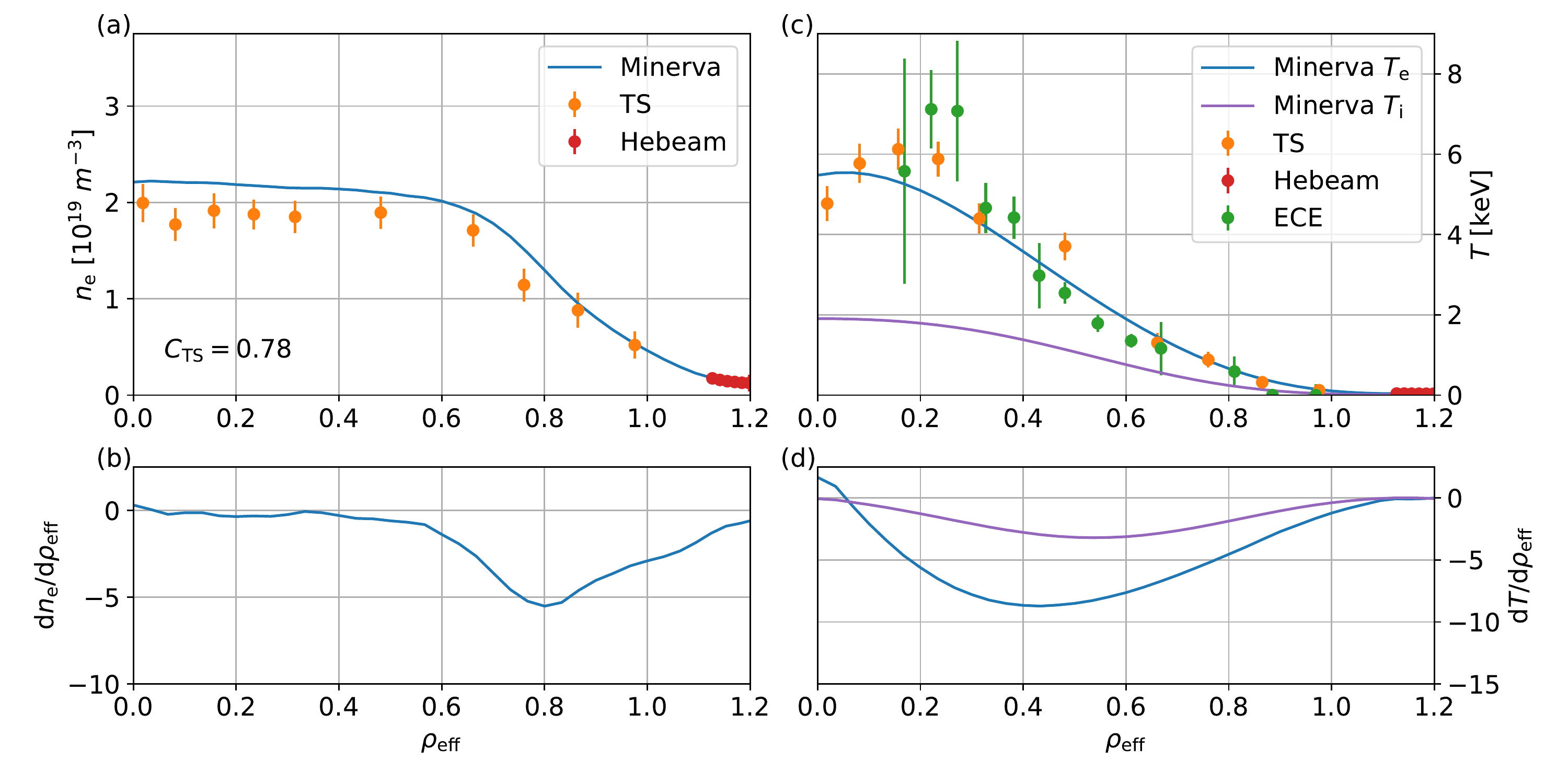}
	\caption{Same as Figure~\ref{fig:profiles} for the inference results of the Bayesian joint model of the interferometer, Thomson scattering, helium beam emission spectroscopy and XICS systems. The ion temperature and $T_\mathrm{i}$ gradient profiles are shown as the purple lines in (c) and (d).}
	\label{fig:full_profiles}
\end{figure}

\begin{figure}
	\centering
	\includegraphics[width=0.9\linewidth]{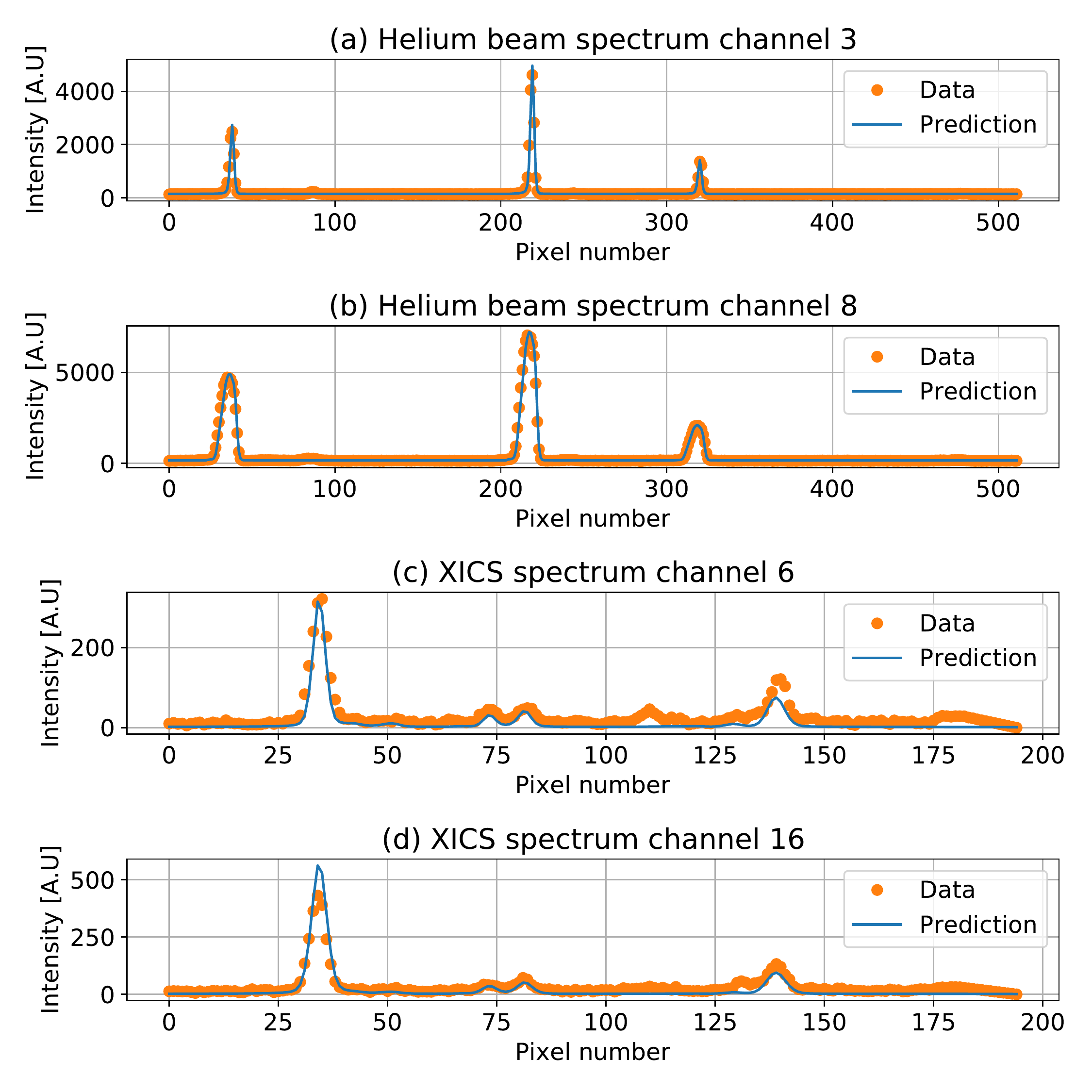}
	\caption{The predictions (in blue) and observations (in orange) of the helium beam spectra of the channel \#3 (near to the divertor) and \#8 (the innermost channel) and the XICS spectra of the channel \#6 (in the edge region) and 16 \#(in the core region) given the profiles shown in Figure~\ref{fig:full_profiles}.}
	\label{fig:predictions_hebeam_xics}
\end{figure}

The maximum \textit{a posteriori} (MAP) solutions of the joint posterior probability of the electron density and temperature, ion temperature profiles are found by the pattern search algorithm \cite{Hooke1961} implemented in Minerva, as shown in Figure~\ref{fig:full_profiles}. The predictions and observations of the helium beam emission and line integrated X-ray spectra are shown in Figure~\ref{fig:predictions_hebeam_xics}. The XICS forward model is substantially complex and computationally expensive, thus full sampling from the joint posterior distribution is left for future work. This can be achieved by a neural network approximation of the XICS Minerva model \cite{Pavone2018}.

Remarkably, we infer these profiles with the optimal values of the hyperparameters (smoothness) and model parameters by maximising the joint posterior probability. A conventional approach to finding the optimal hyperparameters and model parameters is to maximise the posterior probability of these hyperparameters and model parameters, which is proportional to a marginal predictive distribution of the observations, also known as the model evidence. Calculation of the model evidence is computationally challenging because it requires integration over a high dimensional parameter space, therefore this is a major obstacle to apply Bayesian Occam's razor to applications in the real world. On the other hand, calculation of the joint posterior probability does not involve such integration. The joint posterior distribution can be seen as the product of the conditional posterior distribution of the parameters and the posterior distribution of the hyperparameters and model parameters, which can be written as:
\begin{align}
&P\left(n_\mathrm{e},T_\mathrm{e},T_\mathrm{i},\sigma_{n_\mathrm{e}},\sigma_{T_\mathrm{e}},\sigma_{T_\mathrm{i}},\sigma_\mathrm{DI},\sigma_\mathrm{TS},C_\mathrm{TS}|D_\mathrm{DI},D_\mathrm{TS},D_\mathrm{He},D_\mathrm{XICS}\right)\nonumber\\
&=P\left(n_\mathrm{e},T_\mathrm{e},T_\mathrm{i}|\sigma_{n_\mathrm{e}},\sigma_{T_\mathrm{e}},\sigma_{T_\mathrm{i}},\sigma_\mathrm{DI},C_\mathrm{TS},\sigma_\mathrm{TS},D_\mathrm{DI},D_\mathrm{TS},D_\mathrm{He},D_\mathrm{XICS}\right)\nonumber\\
&\times P\left(\sigma_{n_\mathrm{e}},\sigma_{T_\mathrm{e}},\sigma_{T_\mathrm{i}},\sigma_\mathrm{DI},\sigma_\mathrm{TS},C_\mathrm{TS}|D_\mathrm{DI},D_\mathrm{TS},D_\mathrm{He},D_\mathrm{XICS}\right).\label{eq:joint_posterior}
\end{align} 
The joint posterior distribution intrinsically embodies Bayesian Occam's razor through the posterior probability of the hyperparameters and model parameters, and the MAP solution is therefore the optimal profiles with the optimal hyperparameters (smoothness) and model parameters. This does explain the reason why the profiles are not wiggly but optimally smooth in Figure~\ref{fig:full_profiles}.

\section{Conclusions}\label{sec:conclusions}
The Bayesian joint model of the interferometer, Thomson scattering and helium beam emission spectroscopy systems has been developed at Wendelstein 7-X (W7-X). Each of the forward models has been implemented individually and combined together as a joint model in the Minerva framework. The electron density and temperature profiles are given as a function of the effective minor radius and modelled by Gaussian processes with their hyperparameters. The model parameters, for example the calibration factor of the Thomson scattering system, are regarded as additional unknown parameters. The joint posterior distribution of the electron density and temperature profiles, hyperparameters and model parameters is explored by Markov chain Monte Carlo (MCMC) algorithms.

The profile inference has been carried out with different combinations of the three different heterogeneous data sets and virtual observations. The electron density and temperature profiles are inferred with the Bayesian joint model of the interferometer and Thomson scattering system, and the Thomson scattering data is automatically cross-calibrated with the line integrated electron density from the interferometer. In order to exclude physically and empirically improbable solutions, the electron density and temperature are assumed to be not significantly high at the limiter/divertor positions by introducing the virtual observations as a part of the prior distributions. These inferred profiles and calibration factor from the joint posterior distribution with the virtual observations are physically and empirically reasonable and substantially different from those of the joint posterior distribution without the virtual observations due to lack of information of the electron density and temperature in the edge region. Furthermore, in order to compare the inference solutions with the virtual and experimental observations in the edge region, the helium beam emission data is added to the joint model instead of the virtual observations. The profiles inferred with the joint model of the interferometer, Thomson scattering system and helium beam emission spectroscopy systems are reasonable because the helium beam emission data provides the electron density and temperature measurements as well as their measurement uncertainties in the edge region which are crucial to finding the optimal smoothness of the profiles by Bayesian Occam's razor. Nevertheless, when the observations are not sufficiently available, the virtual observations can be a good option to strengthen the model and exclude physically/empirically improbable inference solutions.

We emphasise that these inference solutions have been found with the optimal hyperparameters (smoothness) and model parameters by Bayesian Occam's razor which penalises over-complex models automatically and quantitatively. In other words, these inference solutions neither underfit nor overfit all the measurements. Furthermore, the marginal posterior samples are calculated to obtain the electron density and temperature profiles by taking into account all possible values of the hyperparameters and model parameters given the observations. Remarkably, the joint posterior distribution of the unknown parameters, hyperparameters and model parameters intrinsically embodies Bayesian Occam's razor. The joint posterior probability can be calculated relatively easier than the model evidence, therefore, Bayesian Occam's razor can be applied to the problems in the real world by exploring the joint posterior distribution easier than the model evidence. As shown in this work, the MAP solution of the joint posterior probability distribution given the interferometer, Thomson scattering, helium beam emission spectroscopy and XICS systems provides the electron density and temperature as well as the ion temperature profiles with appropriate model parameters and hyperparameters. Therefore the MAP solution does not either underfit or overfit the data.

\section{Acknowledgement}
This work is supported by National R\&D Program through the National Research Foundation of Korea (NRF) funded by the Ministry of Science and ICT (Grant No. 2017M1A7A1A01015892 and 2017R1C1B2006248) and the the KAI-NEET, KAIST, Korea. This work has been carried out within the framework of the EUROfusion Consortium and has received funding from the Euratom research and training programme 2014-2018 and 2019-2020 under grant agreement No 633053. The views and opinions expressed herein do not necessarily reflect those of the European Commission.

\section*{References}
\bibliography{references}

\end{document}